%% file: main.tex
\documentclass{llncs}
\usepackage[utf8]{inputenc}
\usepackage{graphicx}
\usepackage{hyperref}
\usepackage{algorithm2e}

\begin{document}

\title{Model Checking Cyber-Physical Systems using Particle Swarm Optimization}
\author{Dung Phan\inst{1} 
   \and Scott~A.~Smolka\inst{1}
   \and Radu Grosu\inst{1,2}
   \and Usama Mehmood\inst{1} 
   \and Scott~D.~Stoller\inst{1}
   \and Junxing Yang\inst{1} }
 \institute{Department of Computer Science, Stony Brook University, Stony Brook, NY, USA
\and Department of Computer Science, Vienna Univ.~of Technology, Vienna, Austria}

\maketitle

\input{abstract}
\input{intro}

\input{case-study}
\input{conclusion}

\bibliographystyle{splncs03}
\bibliography{pso-mc}
\end{document}

%% file: abstract.tex
\begin{abstract}

We present a novel approach to the problem of model checking 
cyber-physical systems.  We transform the model checking
problem to an optimization one by designing an objective
function that measures how close a state is to a violation
of a property.  We use particle swarm optimization (PSO) 
to effectively search for a state that minimizes the
objective function.  Such states, if found, are 
counter-examples describing safe states 
from which the system can reach an unsafe state in
one time step.  We illustrate our approach with
a controller for the Quickbot ground rover.  Our PSO
model checker quickly found a bug in the controller
that could cause the rover to collide with an obstacle.

\end{abstract}

%% file: intro.tex
\section{Introduction}
\label{sec:intro}

Dealing with the ``state explosion problem" in model checking has been
an active research area for many years.  Progress has been made by using
abstraction to reduce the size of search space~\cite{Clarke2003counterexample}, 
representing state space symbolically~\cite{McMillan1993symbolic},
bounding the number of steps when unrolling the FSM~\cite{Biere1999verifying},
or using the sheer power of parallel computing~\cite{Bartocci2014towards}.

In this paper, we propose to transform the model checking problem into
an optimization problem.  The intuition is that if we can define an 
objective function that measures how close a state is to an unsafe state,
then we can use available optimization techniques to optimize it.
We propose an objective function such that it has a value of zero if 
the state is one step away from some unsafe state. Otherwise, it has some
positive value depending on how far the next state is from an unsafe state.
This design of objective function makes sure that when an optimizer finds
an optimal state that nullify the objective function, there is a feasible 
transition from that state to an unsafe state.

We use Particle Swarm Optimization~\cite{kennedy95particle} (PSO) to optimize
the objective function.  PSO is a randomized approximation algorithm for 
finding the value of a parameter that minimizes an objective function.  
Using PSO is not mandatory in our approach.  Other optimization techniques
such that gradient descent or simulated annealing can be used in place of PSO.  
We chose PSO because the objective function can be nonlinear and is not
required to be differentiable.  PSO is also proved to be effective in our
case study.  

Our approach differs from statistical model checking in at least two aspects.
First, the goal of statistical model checking is to provide a statistical
guarantee while our goal is to find a bug.  As a result, we terminate the
optimizer as soon as a bug is found.
Second, statistical model checking uniformly explores the state space
whereas in our implementation, although PSO starts from a uniform distribution 
of particles, it moves the particles in strategically calculated directions
so that they converge to an optimal solution.  

A similar approach using genetic algorithms~\cite{godefroid2004exploring} 
was proposed.  However, this approach is more complicated than optimizing
an objective function.

We demonstrate the effectiveness of our approach with a case study of finding
bugs in a controller for the Quickbot ground rover.  Our PSO-based model
checker was able to quickly find a bug that would cause a collision with
an obstacle.

%% file: case-study.tex
\section{The Quickbot Controller Case Study}
\label{sec:case-study}

We conducted a case study of our approach using the Sim.I.am robot simulator.
Sim.I.am allows one to write and test mobile robot controllers in Matlab, 
and then deploy these controllers on actual robots such as the Quickbot~\cite{quickbot14}. 
We model-checked the default Quickbot controller that comes with Sim.I.am
for possible violations of collision-freedom (CF) property. The CF property 
ensures the rover never collides with an obstacle. We used the latest source 
code of Sim.I.am 
(\url{https://github.com/jdelacroix/simiam/tree/cd67b5b97d6781d32333c0a33a51cfd5116640a9}).  

We search for states that would lead to a collision in the next time step.
The state vector is $s = (x, y, \theta, \omega, x_T, y_T)$, where $x, y$ are
initial position of the rover $\theta, \omega$ are initial heading angle
and initial rotational velocity of the rover, $x_T, y_T$ are target location.
We did not include linear velocity $v$ in the state vector because the
Quickbot controller sets $v$ to be a constant and only controls $\omega$.
We design an objective function $J(s)$ that measures how close a state is to
a collision.  For a state $s = (x, y, \theta, \omega, x_T, y_T)$, we initialize
the rover with the initial position $(x, y)$, heading angle $\theta$ and
rotational velocity $\omega$, and then run
the controller for one time step to obtain the next state
$s' = (x', y', \theta', \omega', x_T, y_T)$ (target location does not change).
If the rover collides with an obstacle at the next position $(x', y')$
then $J(s) = 0$.  Otherwise $J(s) = $ min distance from $(x', y')$ to any obstacle. 
Algorithm~\ref{alg:obj-fun} describes the objective function.
Clearly $s$ is a global minimum iff $J(s) = 0$.  We say $s$ is a 
counter-example if $s$ is a global minimum.  
We used Matlab's built-in particle swarm optimization function $particleswarm$
to optimize $J(s)$.  If $particleswarm$ succeeds in finding a global minimum,
then we find a bug in the controller.

\begin{figure}[htb]
  \centering
  \includegraphics[scale=0.25]{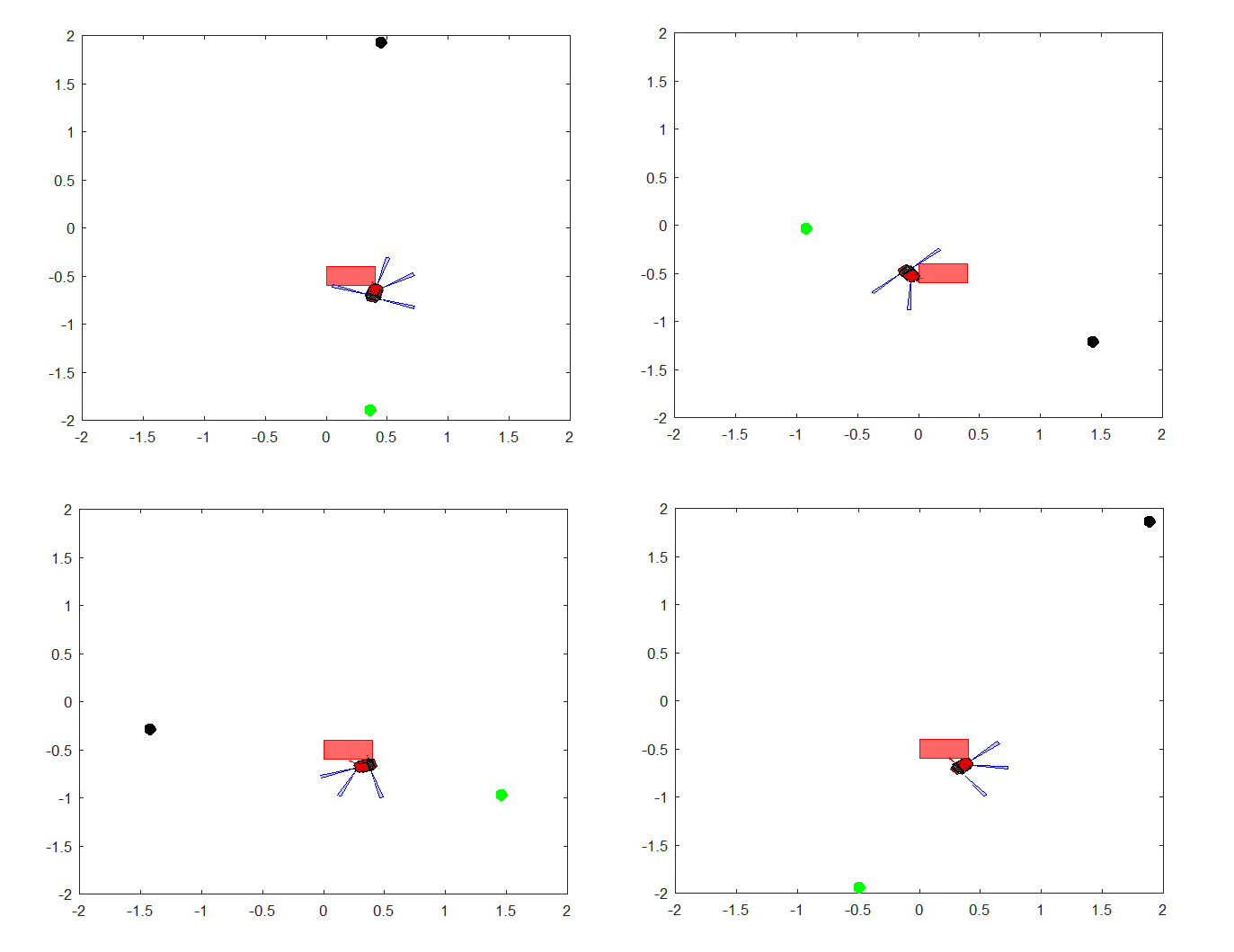}
  \caption{Four counter-examples found from four independent runs 
  of our PSO-based model checking algorithm. The green dots are 
  initial positions, the black dots are targets, and the red
  rectangles are obstacles.}
  \label{fig:counter-examples}
\end{figure}

\begin{figure}[htb]
  \centering
  \includegraphics[scale=0.7]{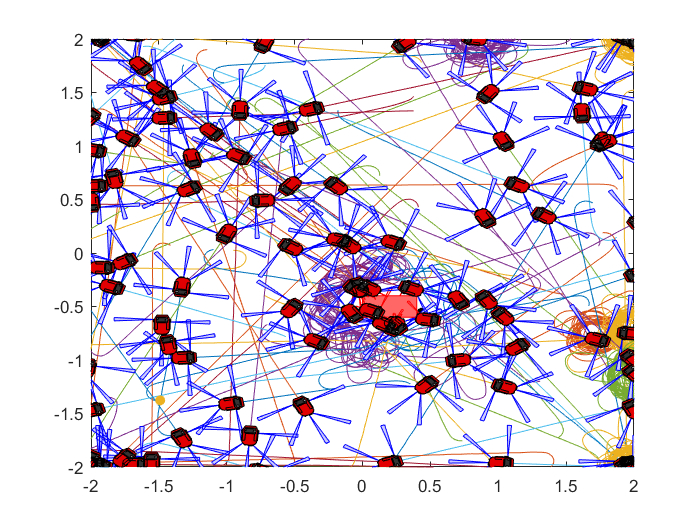}
  \caption{All the paths examined by PSO before it found a counter-example.
  As evidence here, PSO was able to effectively move the particles to converge
  at an optimal solution.}
  \label{fig:paths}
\end{figure}

We ran our PSO-based model checker (PSO-MC) and caught a bug that would make 
the rover collide with an obstacle.  
Each run of the PSO-MC took about 3-4 minutes on a laptop with Core i7-7500 and
16 GB of RAM.  We ran PSO multiple times and each time it found
a different counter-example.  Fig.~\ref{fig:counter-examples} shows four
such counter-examples.  These counter-examples suggest that the Quickbot
controller is susceptible to collision when turning at an obstacle corner.
We plan to further investigate this bug in future work.
Fig.~\ref{fig:paths} shows the states that PSO-MC searched.  
Clearly PSO-MC did not uniformly visit states but instead steer the 
particles to converge at an optimal solution, which is a collision state.

\begin{algorithm}[htb]
 \KwIn{$x, y, \theta, \omega, x_T, y_T$, $dt$, controller, rover, map}
 \KwOut{J}
 \vspace*{0.1in}
 \tcp{Initialize the plant to initial state}
 Init(rover, $x, y, \theta, \omega$)\;
 \tcp{Run controller for one time step}
 Execute(controller, rover, map, $dt$)\;
 \tcp{Get the rover's new position}
 $[x', y']$ = rover.position\;
 \tcp{Calculate cost at the new position}
 if collision($x', y'$, map)\\ 
 \Indp  J = 0\;
 \Indm else\\
 \Indp  J = MinDistanceToObstacle($x', y'$, map)\;
 \Indm end if
 \vspace*{0.1in}
 \caption{Objective function}
 \label{alg:obj-fun}
\end{algorithm}

%% file: conclusion.tex
\section{Conclusions}
\label{sec:conclusions}

We have presented a novel approach based on particle swarm optimization
to model check cyber-physical systems.  We demonstrated how our
approach was able to find a bug in a controller for the Quickbot
ground rover.  Currently the Quickbot case study assumes a map
of obstacle locations and shapes is given.  We plan to include 
obstacles in the search space so that PSO can also find 
obstacle configurations that expose bugs in the controller.  
We also plan to apply our method to other systems and properties.